\documentclass[aps,prl,reprint,superscriptaddress]{revtex4-1}

\usepackage{amsmath,amsfonts,amssymb}
\usepackage{enumitem}
\setlist[enumerate]{itemsep=0mm}

\usepackage[pdftex]{graphicx}
\usepackage{gensymb}
\usepackage{microtype}
\usepackage{bm}\let\vec\bm
\usepackage[normalem]{ulem}

\usepackage[svgnames]{xcolor}
\usepackage[
	colorlinks=True,linkcolor=DarkRed,citecolor=ForestGreen,urlcolor=MediumBlue,
	pdfstartview=FitH,bookmarks=False,pdfpagemode=UseNone
]{hyperref}

\definecolor{green}{rgb}{0.0, 0.7, 0.0}

\newcommand{\Nb}{NbSe$_3$}
\newcommand{\EF}{$E_{\mathrm{F}}$}
\newcommand{\Ga}{$\overline{\Gamma}$}
\newcommand{\Z}{$\overline{\mathrm{Z}}$}

\begin{document}

\title{\boldmath Role of a higher dimensional environment in stabilizing charge density waves in quasi-1D \Nb{} probed by micro-laser angle-resolved photoemission spectroscopy}

\author{C. W. Nicholson}
\email{christopher.nicholson@unifr.ch}
\affiliation{University of Fribourg, Chemin du Mus\'ee 3, 1700 Fribourg, Switzerland}
\author{E. F. Schwier}
\affiliation{Hiroshima Synchrotron Radiation Centre, 2-313 Kagamiyama, Higashi-Hiroshima 739-0046, Japan}
\author{K. Shimada}
\affiliation{Hiroshima Synchrotron Radiation Centre, 2-313 Kagamiyama, Higashi-Hiroshima 739-0046, Japan}
\author{H. Berger}
\affiliation{Ecole Polytechnique F\'ed\'erale de Lausanne (EPFL), 1015 Lausanne, Switzerland}
\author{M. Hoesch}
\affiliation{Deutsches Elektronen-Synchrotron, Notekestrasse 85, 22607 Hamburg, Germany}
\author{C. Berthod}
\affiliation{University of Geneva, 24 quai Ernest-Ansermet, 1211 Geneva, Switzerland}
\author{C. Monney}
\affiliation{University of Fribourg, Chemin du Mus\'ee 3, 1700 Fribourg, Switzerland}

\date{\today}

\begin{abstract}

Utilizing a high energy resolution micro-focused laser with a photon energy of 6.3~eV we investigate charge density waves (CDWs) in the archetypal quasi-1D material \Nb{} using angle-resolved photoemission spectroscopy (ARPES). Doing so allows an exceptionally detailed view into the electronic structure of this complex multi-band system and reveals unambiguously the presence of CDW gaps at the Fermi level (\EF{}). By employing a tight-binding model of the electronic structure, we reveal that the formation of these gaps results from inter-band coupling between electronic states that density functional theory (DFT) finds reside predominantly on separate 1D chains within the material. Two such localized states are found to couple to an electronic state that extends across multiple 1D chains, highlighting the importance of a higher-dimensional environment in stabilizing the CDW ordering in this material. In addition, by investigating the temperature evolution of intra-chain gaps caused by the CDW periodicities far below \EF{} deviate from the behavior expected for a Peierls-like mechanism driven by Fermi nesting; the upper and lower bands of the re-normalized CDW dispersions maintain a fixed peak-to-peak distance while the gaps are gradually removed at higher temperatures. This points towards the gradual loss of long-range phase coherence as the dominant effect in reducing the CDW order parameter which may correspond to the loss of coherence between the coupled chains. Furthermore, one of the gaps is observed above both the known bulk and surface CDW transition temperatures, implying the persistence of short-range incoherent CDW order. Such considerations point to the relevance of a higher-dimensional environment in stabilizing ordered phases in a range of low-dimensional systems.

\end{abstract}

\maketitle

\section{Introduction}

Investigating the emergence of ordered phases such as superconductivity, magnetism and charge density waves (CDWs) is fundamental to furthering our understanding of microscopic interactions in condensed matter. In particular, it is hoped that studying competing or co-existing orders will allow the elucidation of the roles played by different degrees of freedom in driving such exotic phenomena. Prominent examples are charge ordering \cite{Chang2012, Ghiringhelli2012, Peng2018} and short-range charge or magnetic order \cite{Comin2014, SilvaNeto2015, SilvaNeto2016} close to superconductivity in the cuprates and related compounds. An intriguing piece of the puzzle is that many materials exhibiting ordered phenomena are electronically low-dimensional, whereby the electronic wavefunctions are strongly confined to exist in a plane (2D) or within a chain (1D). In such reduced-dimensional systems, a relevant though often unexplored question is to what extent the coupling to a higher-dimensional environment affects their inherent interactions and the emergence of ordered phenomena \cite{Valla2002}. Recent results have shown the importance of inter-plane Coulomb coupling for charge degrees of freedom in an otherwise 2D cuprate \cite{Hepting2018} and that charge correlations can vary from 3D to 2D across a phase transition as a function of doping \cite{Kang2019}. Dimensional considerations have also been highlighted in the CDW mechanism of NbSe$_2$ \cite{Xi2015, Ugeda2016, Weber2018}. Given the current interest in producing heterostructures of 2D-transition metal dichalcogenides, as well as the possibility of investigating quasi-1D edge effects in such materials \cite{Ma2017}, the relevance of mixed dimensionality systems with weak coupling between chains or layers is clear. Here we investigate \Nb{}, a paradigmatic quasi-1D CDW material, in order to investigate the influence of a higher-dimensional environment on charge ordering within a reduced-dimensional system.

\begin{figure*}
\includegraphics[width=1.8\columnwidth]{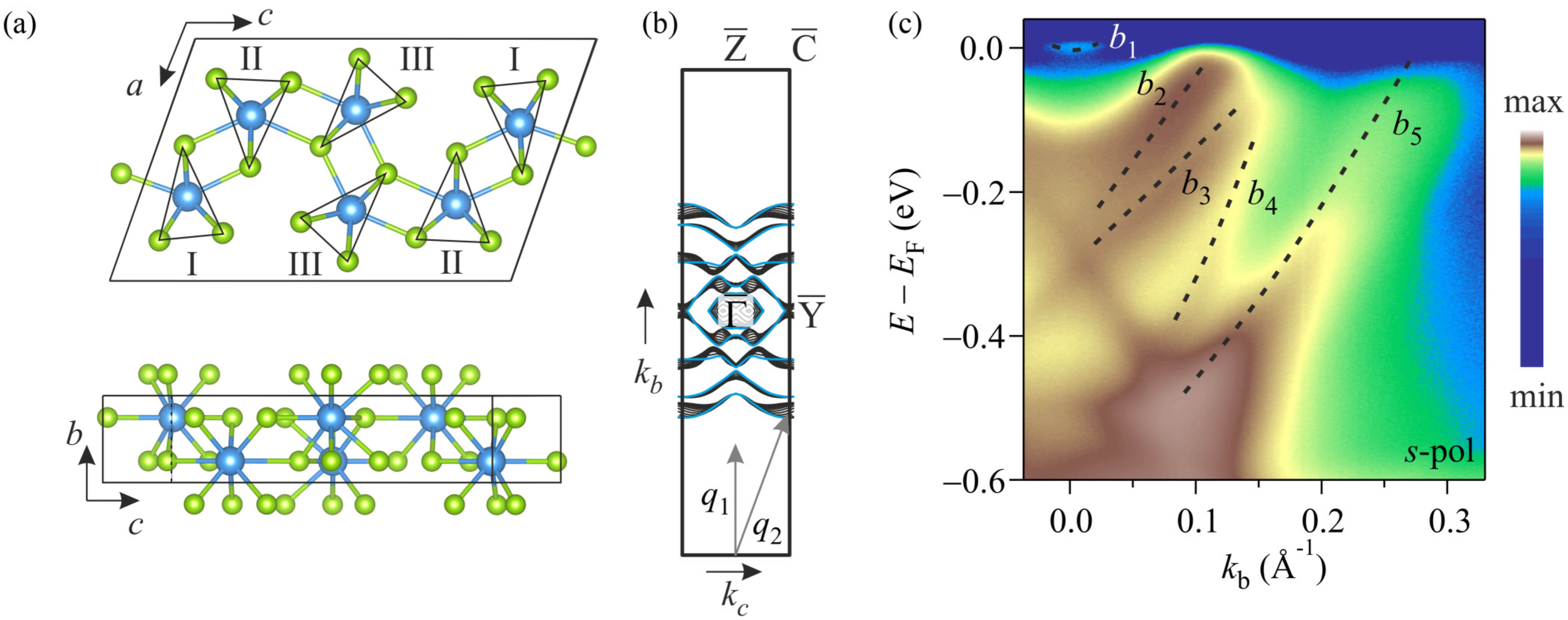}
\caption{a) Atomic structure of \Nb{} in the $ac$ plane (upper) and $bc$ plane (lower). The different chains are emphasized by the marked triangles and are numbered according to the Se-Se distances at the base of the triangle i.e. chain I: 2.49~\AA; chain II: 2.91~\AA; chain III: 2.37~\AA~\cite{Hodeau1978}. Weaker bonds between chains are also shown. b) DFT calculated Fermi surface and surface Brillouin zone. The $k_z$~=~0 cuts are highlighted in blue. The magnitude of the q1 and q2 wave vectors in the bc plane are represented as gray arrows. c) Overview of ARPES data acquired close to the \Ga{}--\Z{} direction with $s$-polarized light showing the 5 expected bands centred at \Ga{}. Data are presented on a log-color scale; dashed lines are a guide to the eye.}
\label{fig:overview}
\end{figure*}

\Nb{} undergoes CDW transitions at $T_1=145$~K and $T_2=59$~K \cite{Monceau1977} with incommensurate modulation wave vectors $\vec{q}_1=(0,0.243,0)$ and $\vec{q}_2=(0.5,0.263,0.5)$ respectively, in units of the reciprocal lattice parameters ($a^{*},b^{*},c^{*}$) \cite{Fleming1978, Hodeau1978}, as drawn in Fig. 1b). The unit cell of \Nb{} contains 3 pairs of inequivalent triangular prism chains, characterized by the Se-Se bond length which forms the base of the triangle in the $ac$ plane, see Fig.~\ref{fig:overview}~a). Nuclear magnetic resonance \cite{Devreux1982, Ross1986} and x-ray diffraction \cite{VanSmaalen1992} measurements have found that the $\vec{q}_1$ CDW predominantly resides on the Nb atoms of type III chains, while the $\vec{q}_2$ CDW exists mainly on chains of type I. This was supported by scanning tunneling microscopy (STM) measurements, with the additional observations of a weak contribution of the $\vec{q}_1$ CDW on chain II and unexpectedly strong contribution of $\vec{q}_2$ CDW on chain III \cite{Brun2009}. This raises the question of to what extent the different chains interact with each other, a possibility also suggested by x-ray diffraction \cite{VanSmaalen1992}. In fact, it has been proposed that the inter-chain Coulomb coupling may facilitate CDW formation \cite{Canadell1990}. These results raise questions regarding the strength of the interactions between these chains, and significance of such interactions in CDW formation.

Early ARPES measurements suggested a Peierls mechanism for the CDWs resulting in gap opening at \EF{} due to nesting between different bands at specific points on the quasi-1D Fermi surface \cite{Schafer2001, Schafer2003b}. However an unambiguous confirmation of CDW gaps at the Fermi level has proved challenging \cite{Nicholson2017, Valbuena2019}, in part due to the difficulty of separating multiple overlapping bands near \EF{}, and the power-law like depletion of spectral weight observed in this system, which thereby obscures the low energy dispersions. Recent ARPES work has suggested the existence of a novel 1D order \cite{Valbuena2019} implying an origin of the CDWs beyond Fermi surface nesting. Indeed, a number of observations seems at odd with a typical Peierls nesting scenario, including the lack of a Kohn anomaly \cite{Requardt2002}, a non-mean field temperature dependence of the $q_1$ gap \cite{Haifeng1999}, and the power-law dependence of spectral weight close to \EF{} \cite{Nicholson2017}. Indeed, despite the clear real-space anisotropy inherent to \Nb{} single crystals, which grow as micrometre hair-like needles, x-ray diffraction \cite{Hodeau1978, Moudden1990}, scanning tunnelling microscopy (STM) \cite{Brun2010} and ARPES \cite{Nicholson2017} have all shown that \Nb{} develops higher-dimensional coherence at low temperatures, suggesting that a simple 1D description does not suffice to explain the observed CDW behavior. This suggests an alternative scenario to Fermi surface nesting may be the driving force for CDW formation. One possibility would be a strong coupling between the electronic and lattice sub-systems \cite{Aubry1992} as discussed in relation to other quasi-1D systems \cite{Lorenzo1998}, that could result in a charge transfer and a change in the bonding conditions as a function of temperature \cite{Wilson1979}. This demands further investigation with ultra-high energy resolution, high-statistics ARPES measurements in the region near \EF{}, with a spot size small enough to probe the small crystals of this material and to avoid a smearing of the momentum resolution. 

In this article, we employ micro-laser ARPES at 6.3~eV with a total energy broadening of less than 2.5~meV and a micro-focused spot of less than 5~$\mu$m in order to determine in detail the electronic structure of \Nb{} close to \EF{}. We reveal unequivocally the existence of gaps at \EF{} related to CDW formation, and confirm the existence of gaps caused by the same CDWs at lower energies as observed previously \cite{Nicholson2017}. A tight-binding simulation of the electronic structure shows that the gaps at \EF{} result from inter-band coupling between states that are predominantly associated with different chains in the real-space unit cell, in contrast to those at lower energies which are caused by intra-band coupling. DFT calculations reveal that most bands have electronic density localized on separated chains in the real-space unit cell. However, one particular state has electronic density on multiple chains and and coupling to more localized states results in the CDW gaps. This clearly supports a more 3D nature of the electronic wavefunctions at low temperatures, and provides a microscopic picture of the CDW. In addition, we find that with increasing temperature the magnitude of the gaps caused by the CDW at lower energies remain constant, in contrast to the expectations for a Peierls system. The upper and lower renormalized bands gradually broaden and merge, which points to a gradual loss of long-range phase coherence as the mechanism removing the order parameter. The fact that the dimensionality of charge excitations is known to change in this material as a function of temperature \cite{Hodeau1978, Nicholson2017, Moudden1990, Brun2010} further hints that the relevant coherence may be between the different chains, highlighting the importance of the role played by higher-dimensional coupling in stabilizing charge order.

\section{Methods}

Single crystals of \Nb{} of typical dimensions $20\times500~\mu$m$^{2}$ were prepared by scotch tape cleaving in ultra high vacuum. ARPES measurements were carried out at the $\mu$-laser ARPES beam line of the Hiroshima Synchrotron Radiation Centre (HiSOR) \cite{Iwasawa2017} over a temperature range 26--180~K. Photons with energy 6.3~eV (197~nm) were generated using a mode-locked Ti:Sapphire laser at 80~MHz repetition rate with 10 picosecond pulse duration to drive frequency addition in non-linear optical crystals. An angular resolution better than 0.05\degree~and total energy broadening less than 2.5 meV were used. The light polarization could be set to either $p$- or $s$-configurations via a zero order wave-plate. The laser was focused on to the sample with a spot size of less than 5~$\mu m$, which is sufficiently small to resolve individual domains. The workfunction of the sample was measured to be 5.0~eV. 

A tight-binding model was constructed to aid the interpretation of the low-energy electronic structure. The model focuses on the key ingredients required in order to reproduce qualitatively the experimental intra- and inter-band CDW gaps. We adopted a two-dimensional three-orbital model for the bands $b_2$, $b_3$, and $b_5$, parametrized by orbital-dependent on-site energies and nearest-neighbor hopping amplitudes along the $b$ and $c$ axes. The Hamiltonian for the CDW was built by considering two cosine potentials with wavevectors $q_1$ and $q_2$ along the $b$ axis and studying the matrix elements in the basis of Wannier functions. For slowly-varying potentials, the leading matrix elements are diagonal in the band index and open the intra-band gaps, while subsequent terms are dipolar and mostly off-diagonal (they involve the gradient of the CDW potential) and open the inter-band gaps. While the diagonal terms were considered in our previous work \cite{Nicholson2017}, the off-diagonal ones were not previously considered. Here, we set them using three parameters $x_{23}$, $x_{25}$, and $x_{35}$ having the unit of a length. Full details of the model and all its parameters are given in the Appendix.

The calculation of the Fermi surface restricted Wannier functions within the real space unit-cell was performed using the OpenMX DFT code which uses norm-conserving pseudo-potentials and pseudo-atomic localized basis functions. The ground-state calculation used the PBE-GGA functional and the atomic basis sets for Se and Nb were set as s3p3d2f1 and s3p3d3f1, respectively. The energy cut-off was 350~Ry and the Brillouin zone was sampled with a 6 x 24 x 4 $k$-grid. Both lattice constants and fractional atomic coordinates were set to the experimental values of the non-CDW unit cell \cite{Hodeau1978}. The highest occupied atomic orbitals (HOMOs) and hence the real space charge distribution, $u_{k,n}(r)$, were calculated across the Fermi surface at various points in the $k_b$ and $k_c$ plane for each of the 5 bands (n = 1-5). Along $k_b$ steps of 0.001 of the $\Gamma-\Gamma$ distance were chosen and the $k_c$-direction was sampled with steps of 0.01 $\Gamma-\Gamma$. In order to restrict the calculation to the region of the Fermi surface, the contribution of each HOMO to the final charge density was damped with a Gaussian profile [$\exp(-(E_B/7.5~meV)^2)$] and any HOMO contribution at binding energies larger than 20~meV was neglected. By taking $\left|\sum_k u_{k,n}(r) \right|^2$ across the Fermi surface for each individual band index, $n$, we arrive at the electronic density corresponding to states at the Fermi level.

\begin{figure}
\includegraphics[width=\columnwidth]{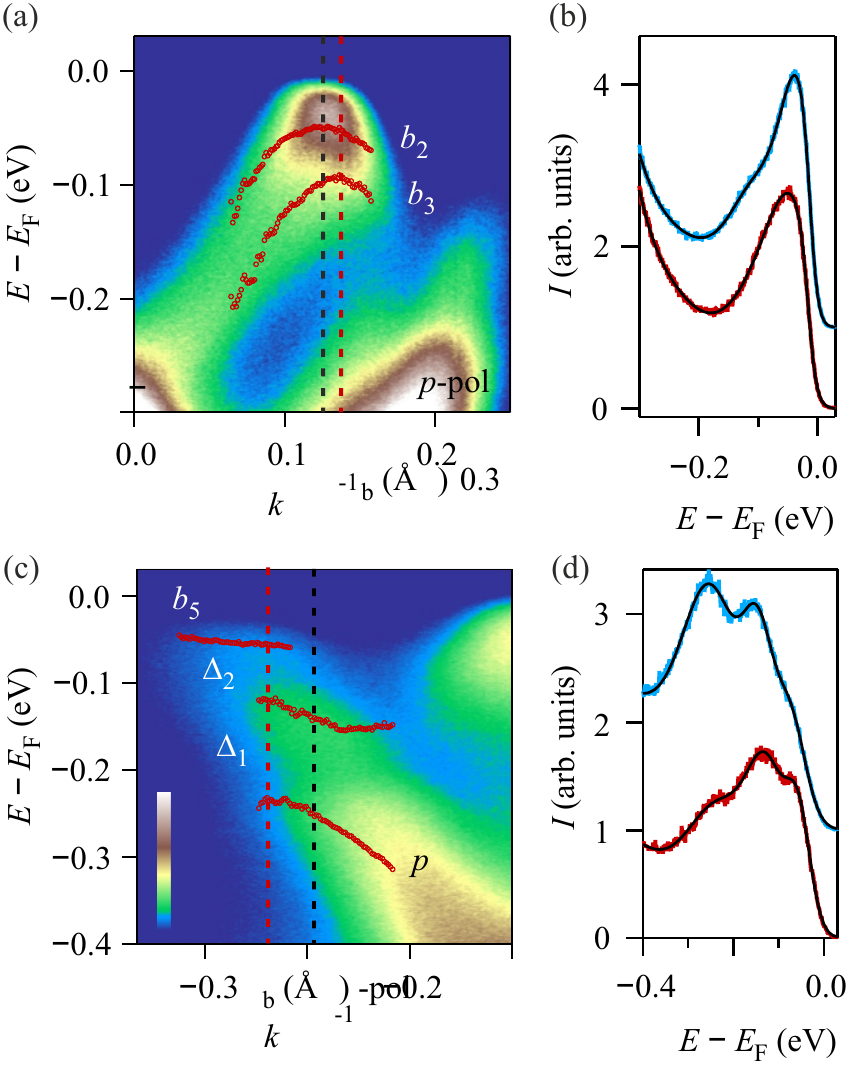}
\caption{a) Zoom on bands $b_2$ and $b_3$ close to \EF{} with $p$-polarized light (linear color scale). EDC fitting results are overlaid to highlight the dispersion. b) Individual EDCs taken at the $k$-position of the band apexes (0.12~\AA$^{-1}$, blue and 0.14~\AA$^{-1}$, red) revealing the CDW gaps. c) Zoom on $b_5$ at higher momenta with $p$-polarized light showing the gaps caused by intra-band coupling as previously observed in \cite{Nicholson2017}. d) Individual EDCs taken at $-$0.25 \AA$^{-1}$ (blue) and $-$0.27 \AA$^{-1}$ (red) highlighting the peaks of the normalized dispersions. All data acquired at 26~K.}
\label{fig:gaps}
\end{figure}

\begin{figure}
\includegraphics[width=\columnwidth]{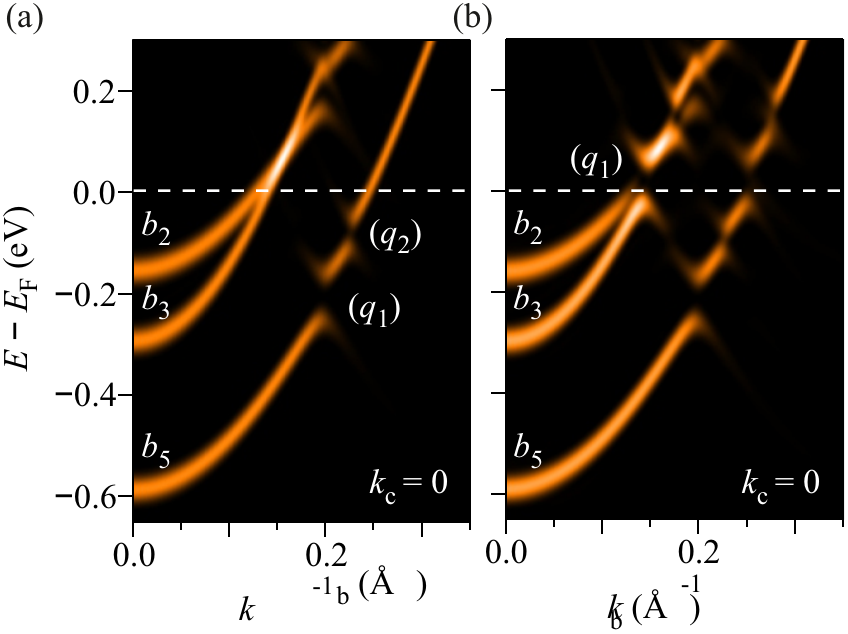}
\caption{Simulated spectral function including CDW periodicities within the tight-binding model described in the text including a) only intra-band coupling and b) both intra- and inter-band matrix elements. This reveals the gaps at \EF{} in $b_2$ and $b_3$ are caused by the respective inter-band coupling between $b_2$ and $b_3$ with $b_5$. The gaps in $b_5$ are confirmed as resulting from intra-band coupling due to the back-folded bands caused by the $q_1$ and $q_2$ periodicities. The wave vectors of the CDWs responsible for opening the gaps discussed in the text are labeled in brackets. Plots with all bare and back folded dispersions overlaid are shown in Fig. 9.}
\label{fig:simulation}
\end{figure}

\section{Results and discussion}

\subsection{Low-energy electronic structure}

ARPES measurements obtained with the $\mu$-focus laser close to the \Ga{}--\Z{} direction in the surface Brillouin zone are presented in Fig.~\ref{fig:overview}c). This reveals the 5 bands predicted by DFT \cite{Schafer2001} and resolved previously by ARPES \cite{Nicholson2017}. The data are presented on a logarithmic color scale due to the high intensity of the $b_5$ band at lower energies compared with states close to \EF{}. The separation of bands and their individual dispersions in the raw data are considerably clearer compared with previous studies, particularly in the region of bands $b_2$ and $b_3$. The observation of the small $b_1$ pocket is in agreement with the STM measurements of Ref.~\cite{Brun2009}, which explained the observed bias voltage dependence by postulating a $b_1$ band with a very shallow minimum below \EF{}. We note that no such pocket was observed in Ref.~\cite{Valbuena2019}. In the region of $b_2$ there additionally appears to be an unexpected sixth band slightly closer to \EF{}. In a material such as \Nb{} which has been considered as a candidate Tomonaga-Luttinger-liquid such an observation prompts speculation of spin and charge separated bands. However a previous investigation suggested that true 1D behavior in this material occurs only at high temperatures \cite{Nicholson2017}. A simpler explanation is that this band is a result of the $k_z$ broadening of the $b_1$ and $b_2$ bands i.e initial to final state transitions are satisfied at multiple points in the bulk Brillouin zone in the out-of-plane direction. DFT predicts the band botto of $b_2$ to disperse by 60~meV between $-0.14$~eV and $-0.20$~eV binding energy over the bulk Brillouin zone in the out-of-plane $k_z$ direction. This matches reasonably well to the experimentally observed difference of 80~meV between $-0.15$~eV and $-0.23$~eV extracted from Fig.~\ref{fig:overview} at the \Ga{}-point. 

To obtain a more detailed view on the electronic states we focus on the two regions presented in Fig.~\ref{fig:gaps}~a) and c), which highlight the dispersions of $b_2$/$b_3$ and $b_5$ respectively. Fitting of the energy distribution curves (EDCs), representative results of which are shown in Fig.~\ref{fig:gaps}~b) and d), allows the extraction of the band dispersions. In the case of bands $b_2$ and $b_3$, maxima are reached at  0.125~\AA$^{-1}$ and 0.135~\AA$^{-1}$ respectively, after which the bands turn towards higher binding energy, leaving gaps of 50~meV and 95~meV ($\pm$~5~meV) between the band maximum and \EF{}. We note that a similar back-folded dispersion can be discerned in $b_2$ in Fig.~\ref{fig:overview}~c), although less clearly due to the log-color scale. Unambiguously observing CDW gaps at \EF{} in \Nb{} has proved challenging due to the multiple overlapping bands and the necessity of very small spot sizes to measure the tiny needle-like crystals. Previous studies that found evidence of gaps at \EF{} were unable to fully disentangle the dispersions of bands $b_2$ and $b_3$ \cite{Schafer2001, Schafer2003b, Nicholson2017} thereby making a precise determination of the gap size difficult. Nevertheless, our current determination of the gap size is in good agreement with both previous STM \cite{Dai1992} and ARPES measurements \cite{Schafer2003b, Nicholson2017}. The dispersion in band-$b_5$ [Fig.~\ref{fig:gaps}~c)] confirms the previous observations of gaps resulting from the CDW periodicities below \EF{} at around $-0.2$~eV \cite{Schafer2001, Schafer2003b, Nicholson2017} and $-0.1$~eV \cite{Nicholson2017} resulting from the overlap of the back-folded $b_5$ band with the bare $b_5$ at positions determined by the $q_1$ and $q_2$ periodicities respectively. In the top right corner of Fig.~\ref{fig:gaps}~c) one can again see the edge of the back-folded intensity of the $b_2$/$b_3$ dispersions. An important observation is the similarity in dispersions obtained with $s$-polarized light in Fig.~\ref{fig:overview}~c) compared with the $p$-polarized data in Fig.~\ref{fig:gaps}~a) and c), in contrast to the data of Ref.~\cite{Valbuena2019} where polarization-dependent matrix elements were proposed to explain the observed differences between $s$- and $p$-polarized light. 

\begin{figure*}
\includegraphics[width=2\columnwidth]{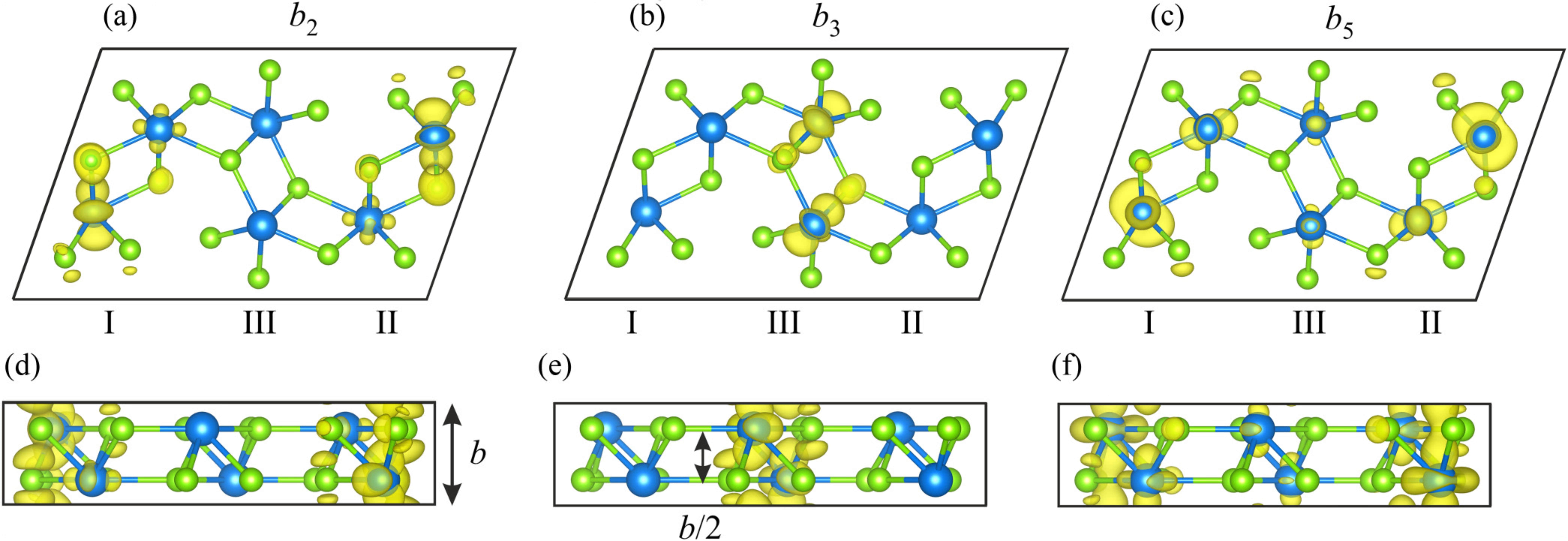}
\caption{a)-c) DFT calculated spatial distribution of electronic density in the $ac$-plane corresponding to states at the Fermi surface in bands a) $b_2$, b) $b_3$ and c) $b_5$. c)-e) The same calculated distributions shown in the $bc$-plane. Examples of electronic density separated by a distance of $b/2$ along the $b$-axis is highlighted by arrows and the relevant coupling index. Such density can contribute to inter-band matrix elements via the dipole operator.}
\label{fig:densityCalc}
\end{figure*}

\subsection{Tight-binding model and DFT}

To determine the origin of the various CDW gaps in the band structure we use a tight-binding model, the results of which are presented in Fig.~\ref{fig:simulation}; full details are given in the Appendix. In the model we consider only bands $b_2$, $b_3$ and $b_5$ as all other bands have little or no spectral weight in the region considered. We first consider only intra-band coupling, which couples the original bands to their own CDW replicas and allows gaps to open. Previously \cite{Nicholson2017}, and here in Fig.~\ref{fig:simulation}~a), the gaps in $b_5$ indeed appear at positions where the original $b_5$ overlaps with its replicas centred at the $q_1$ and $q_2$ wavevectors. Conversely, no gaps appear in either $b_2$ or $b_3$ near \EF{}, as overlap with their own CDW replicas occurs only well above \EF{}. The situation changes, however, once inter-band coupling is is included in the model. Fig.~\ref{fig:simulation}~b) shows the case with $x_{23}$~=~0 and $x_{25}$~=~$x_{35}$~=~0.5$b$, where $b$ is the lattice constant along the chains and $x_{nm}$ sets the matrix element between states $n$ and $m$. In this case, both bands $b_2$ and $b_3$ become gapped at \EF{} as a result of the overlap between their original dispersions with the $b_5$ band back-folded by $q_1$. This extends the description given in Ref.~\cite{Schafer2003b}. For both $b_2$ and $b_3$ the replica of $b_5$ back-folded by $q_2$ only opens gaps above \EF{}. The model therefore allows us to distinguish between the effects of intra- and inter-band coupling and their respective role in the low temperature electronic structure. It is worth noting that the original proposal for a nesting driven CDW in \Nb{} considered the topography of the calculated Fermi surface in relation to the known CDW wavevectors \cite{Schafer2001}, with the result that nested gaps were expected to form at specific locations on the Fermi surface in different bands. Our model predicts a slightly different scenario which removes the uniqueness of these Fermi surface points and instead implies gaps could open all along the Fermi surface, potentially with differing size as observed in other quasi-1D CDW systems \cite{Ahn2004}.

The finding that inter-band coupling is required to form the CDW gaps is particularly interesting with regard to the dimensionality of quasi-1D \Nb{} once the real space extent of the corresponding electronic wavefunctions is considered. To address this we have used DFT to calculate the electronic density distributions of states at the Fermi level associated to bands $b_{1-5}$. This is achieved by a summation of the calculated Kohn-Sham orbitals at different points across the bulk Fermi surface, and taking the square modulus of the result. We have confirmed that the addition of further $k$-points does not alter the calculated outcome. The results are plotted on top of the atomic structure in Fig.~\ref{fig:densityCalc}~a)-c). Band-$b_3$ is found to be the most strongly localized in real-space, essentially being completely confined to type III chains. Band-$b_2$ on the other hand does not appear at all on type III chains, and is instead limited to chains of types I and II. Most importantly, band-$b_5$ is found to have significant contributions across all three chain types and can therefore mediate inter-band coupling across the three different chains, allowing for the hybridization-like gaps in $b_2$ and $b_3$ to open at \EF{}. In fact, the possibility of a non-zero matrix element between bands occurs via the dipole operator (see Appendix) and is maximized for states that are separated by $b/2$ along the $b$-axis. This relies on the, reasonable, assumption that the distribution of electronic density will be similar to the distribution of the maximally localized Wannier functions, which are the quantities used in the tight-binding model. As can be seen from Fig.~\ref{fig:densityCalc}~d)-e), this condition is met for both the $b_2$-$b_5$ and $b_3$-$b_5$ states, but not for for $b_2$-$b_3$ due to the spatial separation of these states. This therefore justifies the choice of the $x_{nm}$'s in the tight binding model ($x_{23}=0, x_{25}=0.5b, x_{35}=0.5b$). Such an inter-chain mixing of states highlights the higher-dimensional nature of the CDW phase in \Nb{} at low temperatures, as previously suggested by multiple techniques, and may also explain the observation of strong contributions of different CDW $q$-vectors occurring on different chain types with STM \cite{Brun2009}.

\subsection{Temperature dependence of gaps}

In the classic Peierls picture both the electronic gap opening at \EF{} due to the lattice distortion, as well as long-range phase coherence of this order, develop at the transition temperature. Upon lowering the temperature further, the electronic gap gradually widens at the same time as the atomic displacement increases. Such expectations hold in the limit of weak electron-phonon coupling, but in materials where a strong or $k$-dependent coupling exists the situation is expected to be different \cite{Rossnagel2011}. In particular, there is the possibility that the electronic gaps survive well above the transition temperature while the CDW loses long-range phase coherence. This has previously been observed in various CDW materials \cite{Chaterjee2015, Yokoya2005, Hoesch2019} as well as in superconductors \cite{Kanigel2008}.

\begin{figure}
\includegraphics[width=\columnwidth]{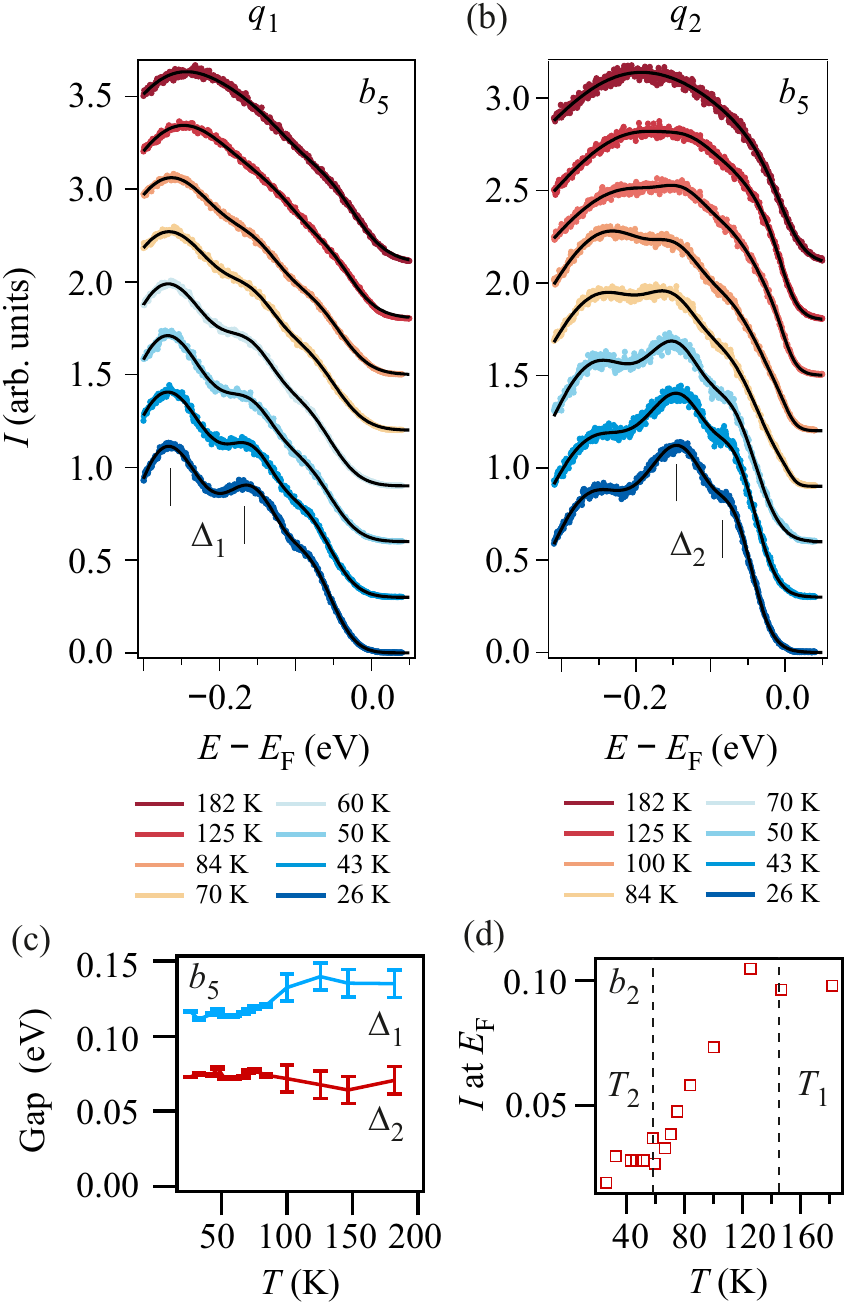}
\caption{Temperature dependence of EDCs corresponding to the $q_1$ CDW in band-$b_5$ as shown in Fig.~\ref{fig:gaps}~c). The EDCs in a) correspond to the lower CDW gap and b) the upper. Curves are offset vertically for clarity. The $\Delta_2$ gap can be seen in the EDCs until at least 84~K, which is above the expected transition temperature. c) Gap size of $q_1$ and $q2$ CDWs in band-$b5$ extracted through fitting of temperature dependent EDCs. The gap size is found to remain constant within error bars as the temperature is increased, but is gradually removed at higher temperatures pointing towards a loss of phase coherence as the mechanism removing the CDW. d) Intensity of band-$b_2$ at \EF{} revealing clear changes in the spectral weight at \EF{} between the two transition temperatures.}
\label{fig:temperature}
\end{figure}

To investigate the possibility of such behavior in \Nb{} we have performed temperature dependent measurements in the range 26--180~K, which encompasses both CDW transition temperatures. We have focused on the evolution of the CDW gaps in band-$b_5$, as in the region of \EF{} where gaps occur the spectral weight develops a power-law like dependence at higher temperatures [Fig. 6(a)] which complicates a systematic analysis. The EDCs presented in Fig.~\ref{fig:temperature}~a) and b) are obtained at the two positions marked in Fig.~\ref{fig:gaps}~c). The peaks correspond to the upper and lower parts of the renormalized dispersions, and therefore reveal the gap size. The fact that three peaks can be seen in each panel reflects the fact that the EDCs at each position also catch a little of the neighboring gap. Upon increasing the temperature the gaps are gradually filled with spectral weight, suggestive of a loss of phase coherence, as supported by x-ray diffraction measurements [see Appendix, Fig.~\ref{fig:diffraction}]. Additionally, the double peak structure of the $\Delta  _2$ gap persists to at least 100~K, well above the bulk transition temperature of the $q_2$ CDW 59~K and even above the surface transition temperature of 70-75~K obtained with STM \cite{Brun2010}. Evidence of short-range CDW order has also been found in X-ray scattering data around 10~K above the $q_1$ \cite{Moudden1990} and $q_2$ transitions, and ARPES measurements at 300~K which were ascribed to 1D fluctuations \cite{Schafer2001a}. In order to probe the gap evolution further, and to obtain the evolution of the gap size as a function of temperature, we fit the EDCs with 3 Gaussians [overlaid in Fig. 5a) and b)] to extract the peak positions and corresponding gap sizes, which are presented in Fig.~\ref{fig:temperature}~c). Within error bars the gap size is constant over the full temperature range in both cases [Fig.~\ref{fig:temperature}~c)] i.e. the distance between peaks does not change. 

We note that fitting at higher temperatures becomes much less well defined due to the broadening of the EDCs. In the region of the gap this is likely a result of a mixtureof effects: broadening due to increased electron-phonon scattering at higher temperatures and the intrinsic removal of the gap. Distinguishing these effects is a delicate matter, and is beyond the scope of the current article. However, the fact that multiple effects may contribute simultaneously does not change the central result: that the relative positions of the upper and lower CDW dispersions, and therefore the CDW gaps, do not change as a function of temperature. Since it is known that the CDWs do not survive up to room temperature, a mechanism is still required to remove the gaps. Our data is therefore suggestive of a loss of phase coherence as the mechanism by which the CDW is removed, rather than it closing in a Peierls-like manner. This hypothesis is supported by temperature dependent X-ray diffraction
measurements (see Appendix A, Fig. 8) which find an increase of intensity in the CDW super-structure peaks, implying increased coherence length at low temperatures.

Curiously, the band positions themselves are not constant, even though the gaps sizes are. This is evidenced by a shift of both upper and lower dispersions by around 30~meV as a function of temperature [Fig. 7a) and b)]. A similar shift is observed in band b3 well away from \EF{}. This rigid band shift of multiple bands is suggestive of a change in the chemical potential as a result of the
CDW gap opening, but may also result from the transfer of charge between Nb atoms on different chains [41]. The exact relation of this shift to CDW formation is unclear at present. This slight shift of the bands may also be partially reflected in the change of spectral weight at \EF{} [Fig. 5d)]. Here one sees the increase of spectral weight with increasing temperature, again over the temperature region defined by the two CDW transitions. However, given the change in line shape at higher temperatures which becomes power-law like, it is difficult to say exactly what mechanism causes the change of spectral weight and what relation this bears to the macroscopic properties of this material.

\subsection{Discussion}

The present observations of CDW gaps with inter-chain character highlight the role that a higher dimensional environment plays in stabilizing the long-range CDW order in \Nb{}. In addition, the fact that the CDW appears to be removed via a loss of phase coherence, as evidenced by the gradual filling of the CDW gaps at constant gap size, suggests a clear picture by which the CDW ordering is influenced by the phase coherence between chains. This leads us to the following picture of how the CDW is removed: At low temperatures, CDW order is well developed and is stabilized by long-range coherent electronic states across the chains in a 3D environment. As the temperature is increased, the coherence of the electronic states reduces, with the result that they become more localized to individual chains, thereby gradually reducing the dimensionality of the system and weakening the inter-chain CDW phase coherence. At higher temperatures the CDW is completely washed out as the electronic states become localized to individual chains due to their very short coherence lengths and therefore the chains completely lose their phase coherence. The high-temperature 1D phase may also host Tomonaga-Luttinger liquid behavior. Such a picture complements the description given in our previous publication where we estimated the effective crossover energy scale from 3D to 1D based on a tight-binding fit of the quasi-1D Fermi surface \cite{Nicholson2017}. This revealed 3D behaviour at low temperatures is in accordance with a number of previous works \cite{Hodeau1978, Moudden1990, Brun2010}, which can be seen in light of the above discussion as resulting from the increase of electronic coherence between chains, and which allows the development of CDW order.

That the CDW order parameter is removed by phase incoherence, rather than amplitude reduction, itself implies a mechanism different to the traditional Peierls model. An alternative description such as the metal-metal bonding approach proposed by Wilson \cite{Wilson1979} may be appropriate, which is analogous to the Jahn-Teller distortion in molecules. In this scenario, a change of orbital occupancy as a function of temperature changes the bonding and anti-bonding populations for particular bonds, resulting in the CDW. Similar conclusions were drawn from the calculations of Ref.~\cite{Canadell1990} which further highlighted the role of inter-chain Coulomb interactions is stabilizing the CDWs. The observation of the rigid band shift during CDW formation presented above may indeed point in the direction of a redistribution of charge in the Nb states at \EF{}, although the details remain unclear. The potential relevance of such a description can also be seen in the context of recent time-resolved experiments on femtosecond timescales where the transient occupation of bonding and anti-bonding states was shown to photo-induce a CDW transition \cite{Nicholson2018a} in another quasi-1D system. The band shift and gradual gapping of the Fermi surface as seen in Fig.~\ref{fig:bandShift} and Fig.~\ref{fig:temperature}~d) certainly suggest a redistribution of charge, which points in the direction of a more chemical-like description of the CDW. On the other hand, the shift of the bands may be explained by a shift of the chemical potential due to the opening of the CDW gaps. An analysis of this is therefore complicated by the power-law depletion that obscures the gap at \EF{}. Further studies in this direction may shed light on the role of particular state occupancy on the CDWs. 

\begin{figure}
\includegraphics[width=\columnwidth]{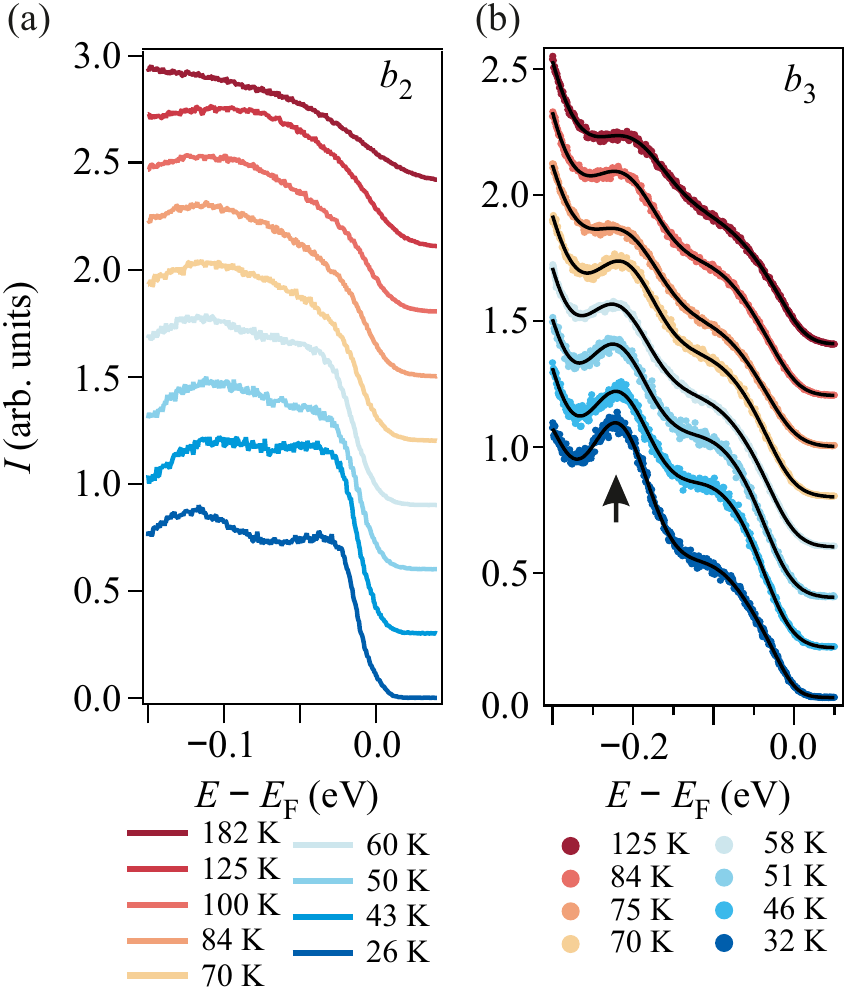}
\caption{a) Temperature dependent EDCs obtained at $k=0.135$~\AA$^{-1}$ revealing the evolution of the CDW gap at \EF{} into the power-law line shape characteristic of the high temperature phase. b) Temperature dependent EDCs of band $b_3$ at $k=0.06$~\AA$^{-1}$ which are fitted to extract the rigid band shift shown in Fig. 7c).]}
\label{fig:temperature_extra}
\end{figure}

Further evidence of a non-Peierls scenario comes from the distribution of spectral weight close to \EF{}, which has the form of a power-law or pseudogap at low temperatures \cite{Nicholson2017}. Such behaviour has been observed in a number of quasi-1D systems, although to date no general explanation has been found for this. Some instances have been attributed to Tomonaga-Luttinger liquid behaviour \cite{Wang2006, Blumenstein2011, Ohtsubo2015, Watson2017}, where all spectral functions are predicted to take a power-law form \cite{Voit1993,Voit1995}, although other signatures of TLL behavior in these materials such as spin-charge separation have proved elusive. In addition, as discussed above, we predict that TLL behavior should only appear at higher temperatures in \Nb{} once the phase coherence of electronic stats is reduced, restricting them to individual chains. It therefore seems unlikely that the particular depletion of spectral weight is a result of the TLL. Disorder has been suggested to play a role in spectral weight depletion in some systems \cite{Starowicz2002}, although this may be less applicable in well-ordered single crystals with very low defect density such as \Nb{}. An alternative scenario that predicts both CDW formation and a spectral shape in photoemission similar to a power law are strong electron-phonon coupling theories, as discussed in Ref.~\cite{Rossnagel2011}. Such models have been applied to both 1D \cite{Lorenzo1998} and surface 2D systems \cite{Tournier-Colletta2010} and may be applicable to \Nb{}. We note that strong coupling theory also predicts commensurate lattice distortions, compared with incommensurate in the weak-coupling limit. It is interesting to consider therefore, that the combined CDWs in \Nb{} is "nearly commensurate" meaning that the combined period of the $q_1$+$q_2$ is close to an integer number of lattice constants which hints at an intermediate strength coupling in \Nb{}. The current observations in \Nb{} therefore suggests the appropriateness of a description where the role of Fermi surface nesting is reduced, and $k$-dependent electron-phonon coupling and inter-chain coupling play central roles \cite{Canadell1990}.

\section{Summary}

In summary, we have investigated the low energy electronic structure of \Nb{} with $\mu$-spot laser-ARPES with very high energy resolution. We have clearly demonstrated gaps due to CDW formation both at and below \EF{}. We have discussed their origin in terms of inter- and intra-band matrix elements respectively, and placed this in the context of the spatial extent of the related electronic wavefunctions, which are predominantly found to reside on individual 1D chains but with coupling mediated by more delocalized states. This highlights the higher-dimensional environment that stabilizes CDW behaviour in this quasi-1D material. In addition, we have followed the positions of the renormalized dispersions as a function of temperature and found that the gap size stays constant, and is even visible above the CDW transition temperatures. Since the fixed gap is gradually filled, we conclude that it is the loss of long-range phase coherence rather than the reduction of the CDW amplitude that removes the order parameter. It is plausible that the loss of coherence relates to the phase coherence between the electronic states on different chains that contribute to the CDW gaps. We have discussed our results in the context of different mechanisms to the standard Peierls nesting scenario. These considerations extend beyond ordered phases to other low dimensional systems such as heterostructures of 2D transition metal dicalchogenides or 1D edge states, where tuning the coupling between layers is a topic of current interest.

\clearpage
\section{Appendix}

\subsection{Extended data}

To show the increasing influence of the power-law spectral shape on the CDW gap at \EF{} as a function of temperature, Fig. 6a) shows the EDCs through the maxima of bands $b_2$ and $b_3$. This complicates the analysis of the spectral weight and the CDW gap. Fig. 6b) shows EDCs from band b3 from which the band shift discussed in the man text (and below) is extracted by fitting.

\begin{figure}[t]
\includegraphics[width=\columnwidth]{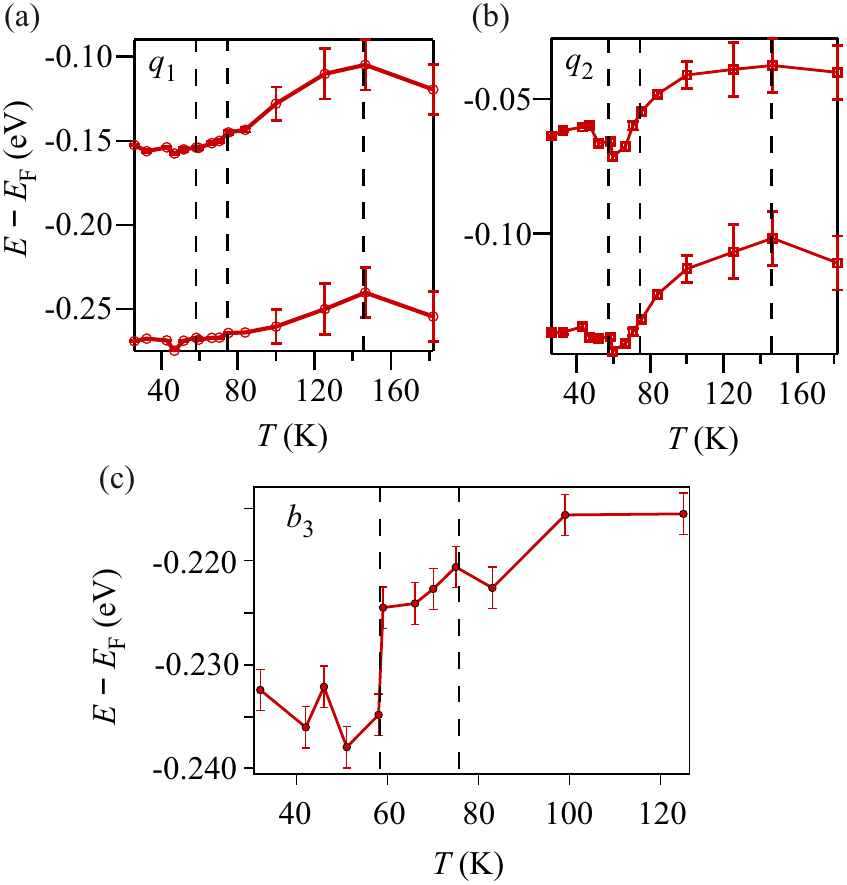}
\caption{Extracted band positions as a function of temperature in band-$b_5$ for a) the $q_1$ and b) the $q_2$ CDWs. All bands are found to shift over a temperature range between the two CDWs transitions. $T_{c2}$, $T_{c2, surface}$ and $T_{c1}$ are marked by dashed lines from left to right. The corresponding raw data and extracted gap sizes are shown in Fig.~\ref{fig:temperature}. c) Extracted position of band $b_3$ from the EDCs in Fig. 6.]
}
\label{fig:bandShift}
\end{figure}

Although the gap sizes ($\Delta_{1,2}$) remain approximately constant as a function of temperature, as shown in Fig.~\ref{fig:temperature}~c) the upper and lower positions of the renormalized dispersion clearly change. In Fig.~\ref{fig:bandShift} the extracted positions for both $q_1$ and $q_2$ related dispersions are shown. It is notable that the shift occurs exactly within the temperature region where the CDW transitions occur. A shift of the same magnitude occurring in band b3 [Fig. 7c)] is suggestive of a charge redistribution as discussed in the main text.

Single crystal X-ray diffraction data were obtained at the Swiss-Norwegian beamline BL01A at the ESRF. The sample was illuminated with monochromatic X-rays of 0.717 \AA wavelength and step-wise rotated with 1\degree oscillations while acquiring the scattered x-rays on an area detector (MAR345 image plate). The samples were cooled in the dry nitrogen gas flow of a cryostream cooler (Oxford Instruments). The indices of diffraction spots were determined and arbitrary planes of X-ray scattering intensity reconstructed using the software Crysalis (see Fig. 7a for a high symmetry plane). The intensity of individual diffraction spots was also determined against the background of adjacent diffuse X-ray scattering and background using Crysalis (Fig. 7b).
The onset of intensity at the incommensurate scattering vector (1, 1.241, 0) [Fig. 8b] is consistent with a periodic lattice distortion (PLD) related to the CDW modulation q1, which sets in sharply at $T_1$ and develops fully towards lower temperatures. This crystallographic behavior of the long-range ordered PLD is consistent with the ARPES observation of gaps becoming better defined with lowering temperature but with fixed gap size.

\subsection{Three-band tight-binding model}

The effect of the CDW on the low-energy dispersion in NbSe$_3$ is most conveniently discussed in the basis of Wannier states $|\vec{R}n\rangle=N^{-1/2}\sum_{\vec{k}}e^{-i\vec{k}\cdot\vec{R}}|\vec{k}n\rangle$, where $|\vec{k}n\rangle$ are the Bloch states. $\vec{R}$ represents any vector of the Bravais lattice, $n$ is the band index, $\vec{k}$ is a wavevector in the first Brillouin zone, and the state $|\vec{R}n\rangle$ is localized around $\vec{R}+\vec{\tau}_n$ with $\vec{\tau}_n$ some vector in the primitive cell. While the entanglement of several bands renders the practical construction of Wannier functions difficult for NbSe$_3$, it is sufficient for our purposes that they are well defined formally. In the absence of CDW, the Hamiltonian $H_0$ is diagonal in the band index and the matrix elements $\langle\vec{R}n|H_0|\vec{R}'n\rangle$ fall off rapidly with increasing $|\vec{R}-\vec{R}'|$. Our two-dimensional tight-binding model retains the local and nearest-neighbor matrix elements in the $(b,c)$ plane, i.e., $\langle\vec{R}n|H_0|\vec{R}n\rangle=-\mu_n$, $\langle\vec{R}n|H_0|\vec{R}'n\rangle=-t_{nb}$ if $|\vec{R}-\vec{R}'|=b$, and $\langle\vec{R}n|H_0|\vec{R}'n\rangle=-t_{nc}$ if $|\vec{R}-\vec{R}'|=c$. The corresponding electronic dispersion is $E_{\vec{k}n}=-2t_{nb}\cos(k_bb)-2t_{nc}\cos(k_cc)-\mu_n$. Restricting to the subspace of bands $b_2$, $b_3$, and $b_5$, we choose the parameters such as to fulfill the following conditions extracted from the data: the hopping amplitude along $c$ is 27~meV \cite{Nicholson2017}; the minima of $b_2$, $b_3$, and $b_5$ are at $-0.15$, $-0.29$, and $-0.58$~eV; the $q_1$ and $q_2$ gaps in $b_5$ are at $-0.2$ and $-0.09$~eV; the crossing between $b_3$ and the $q_1$-folded $b_5$ is at $+20$~meV; the crossing between $b_2$ and the $q_1$-folded $b_5$ is at $+35$~meV. The ensuing parameters are $(\mu_2,t_{2b},t_{2c})=(-1.488,0.792,0.027)$\,eV, $(\mu_3,t_{3b},t_{3c})=(-2.298,1.267,0.027)$\,eV, and $(\mu_5,t_{5b},t_{5c})=(-1.148,0.837,0.027)$\,eV.

\begin{figure}[t]
\includegraphics[width=\columnwidth]{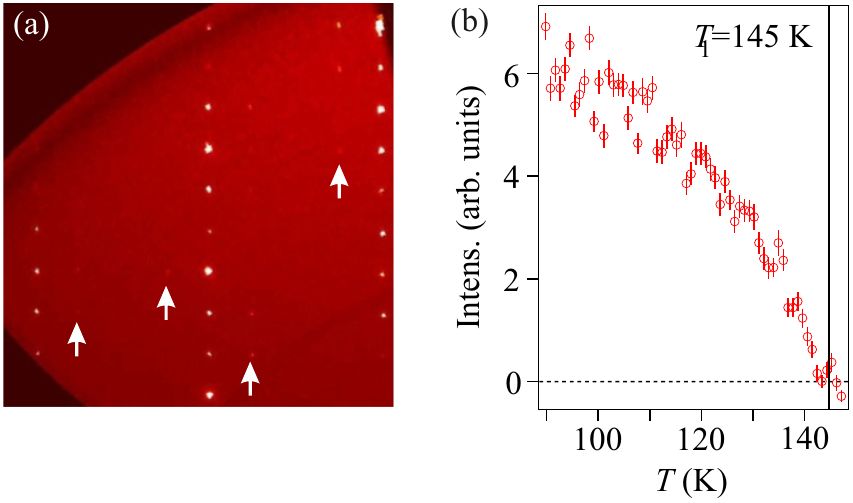}
\caption{a) X-ray diffraction obtained in the (5$kl$) plane at 80~K. Superstructure spots corresponding to the $q_1$ CDW are marked by the white arrows. b) Intensity of the (1 1.241 0) spot, corresponding to the $q_1$-CDW, as a function of temperature. Intensity for this peak first appears at 143~K. The gradual increase of the peak intensity implies the increase of the CDW coherence length towards lower temperatures, consistent with the ARPES observation of gaps becoming better defined but with fixed gap size.
}
\label{fig:diffraction}
\end{figure}

\begin{figure}[t]
\includegraphics[width=\columnwidth]{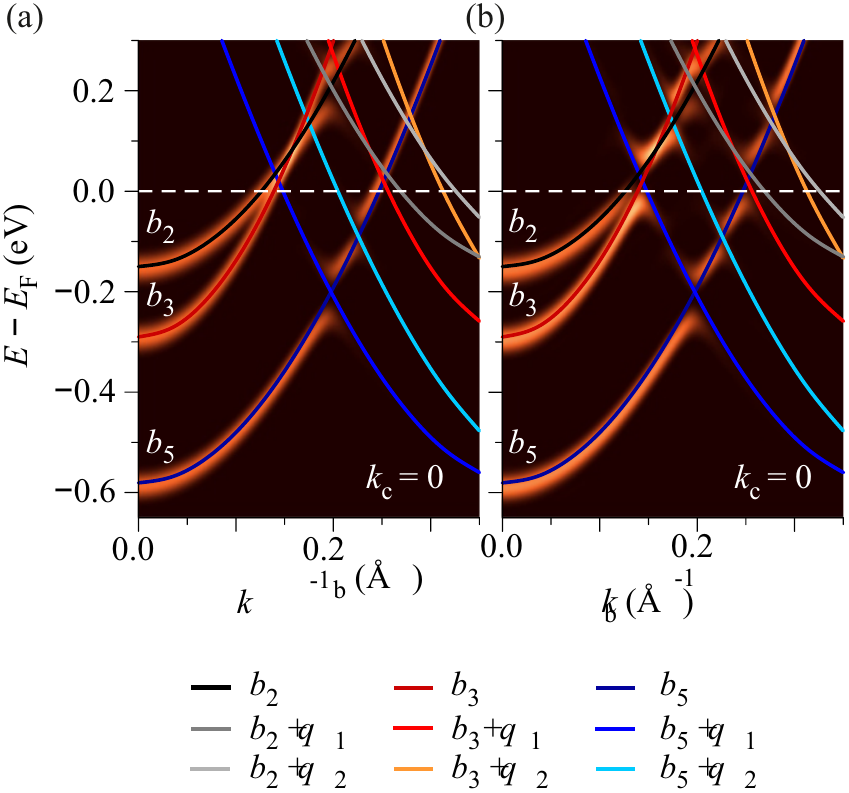}
\caption{Tight-binding calculations for the CDW phase as presented in the main text a) excluding and b) including inter-band coupling with the bare dispersions and those of the back-folded bands overlaid. The bare dispersion from each of the bands are labeled $b_1$, $b_2$ and $b_3$respectively. The back-folded dispersions resulting from the $q_1$ and $q_2$ CDWs are color coded for clarity.
}
\label{fig:modelDetail}
\end{figure}

The CDW acts on the electrons as a local potential $V_{\mathrm{CDW}}(\vec{r})$. The detailed shape of this potential in real space is unknown for NbSe$_3$, except for the fact that is has two components of known periodicities. The simplest Ansatz keeps only the first Fourier component:
\begin{equation}
V_{\mathrm{CDW}}(\vec{r})=2V_1\cos(\vec{q}_1\cdot\vec{r}+\varphi_1)
+2V_2\cos(\vec{q}_2\cdot\vec{r}+\varphi_2).
\end{equation}
For simplicity, we ignore the component of $\vec{q}_2$ normal to the $b$ axis and we ``rationalize'' the periods to the values $q_1b=7\pi/16$ and $q_2b=\pi/2$, such that $V_{\mathrm{CDW}}$ is commensurate with a period $32b$ \cite{Nicholson2017}. The amplitudes $2V_1=95$~meV and $2V_2=65$~meV are taken as the low-temperature values of the $q_1$ and $q_2$ gaps in $b_5$. Like for $H_0$, we represent $V_{\mathrm{CDW}}$ in the Wannier basis and retain the leading contributions. For two states $|\vec{R}n\rangle$ and $|\vec{R}'m\rangle$, $V_{\mathrm{CDW}}(\vec{r})$ may be expanded around the point $\vec{r}_0=\frac{1}{2}\left(\vec{R}+\vec{\tau}_n+\vec{R}'+\vec{\tau}_m\right)$ where the overlap of the Wannier functions is largest. The expansion is meanigful if $V_{\mathrm{CDW}}$ varies slowly in space as compared to the Wannier functions. The matrix elements read
\begin{multline}\label{eq:expansion}
\langle\vec{R}n|V_{\mathrm{CDW}}|\vec{R}'m\rangle
=V_{\mathrm{CDW}}(\vec{r}_0)\langle\vec{R}n|\vec{R}'m\rangle\\
+\vec{\nabla}V_{\mathrm{CDW}}(\vec{r}_0)\cdot
\langle\vec{R}n|\vec{r}-\vec{r}_0|\vec{R}'m\rangle+\ldots
\end{multline}
Owing to the orthogonality of the Wannier states, the first term in the right-hand side of Eq.~(\ref{eq:expansion}) is diagonal and shifts the onsite energies by the orbital-dependent value $V_{\mathrm{CDW}}(\vec{R}+\vec{\tau}_n)$ without mixing the bands. The second term of the expansion has in general both diagonal and off-diagonal contributions in the band indices, that fall off rapidly with increasing $|\vec{R}-\vec{R}'|$. The terms with $|\vec{R}-\vec{R}'|=b$ change the hopping amplitudes $t_{nb}$ by a value modulated in space according to the CDW. As $V_1,V_2\ll t_{nb}$, we neglect these changes and retain intra-cell terms with $\vec{R}=\vec{R}'$. Furthermore, we ignore the diagonal terms proportional to $\langle\vec{R}n|\vec{r}-\vec{r}_0|\vec{R}n\rangle$, as these terms vanish identically for Wannier functions that have inversion symmetry around the point where they are localized. Our minimal tight-binding model for the CDW finally reads
\begin{multline}\label{eq:CDW}
\langle\vec{R}n|V_{\mathrm{CDW}}|\vec{R}'m\rangle\approx
\delta_{\vec{R}\vec{R}'}\big[
\delta_{nm}V_{\mathrm{CDW}}\left(\vec{R}+\vec{\tau}_n\right)\\
+(1-\delta_{nm})\vec{d}_{nm}\cdot\vec{\nabla}V_{\mathrm{CDW}}
\left(\vec{R}+\textstyle\frac{\vec{\tau}_n+\vec{\tau}_m}{2}\right)\big].
\end{multline}
The components of the vectors $\vec{d}_{nm}=\langle\vec{0}n|\vec{r}-(\vec{\tau}_n+\vec{\tau}_m)/2|\vec{0}m\rangle$ along $b$ are the only one required, as the gradient of the CDW potential lies along the $b$ axis. We denote these components $x_{23}$, $x_{25}$, and $x_{35}$.

We compute the spectral function using a Chebyshev expansion as explained in Ref.~\cite{Nicholson2017}. The order of this expansion is such that the energy resolution is $40$~meV, which leads to a broadening of the bands as seen in Fig.~\ref{fig:simulation}. Our calculations show that the spectral function is insensitive to the phases $\varphi_1$ and $\varphi_2$, which we therefore set to zero. Likewise, the precise choice of the positions $\vec{\tau}_n$ only amounts to phase shifting the potential and does not influence the results.

\clearpage
\section{Acknowledgments}
The laser ARPES measurements at HiSOR were performed with the approval of the Proposal Assessing Committee (Proposal Numbers: 16BU011). Liquid Helium forthe ARPES measurements was supplied by the N-BARD, Hiroshima University. We thank Alexei Bosak, Dmitry Chernyshov, and Phil Pattison for suggesting the X-ray diffraction data acquisitions shown in Appendix A and for help with this experiment. This project was supported by the Swiss National Science Foundation (SNSF) Grant No. P00P2 170597.

%

\end{document}